\documentclass[11pt]{JHEP3}

\newcommand{\dalpha}{{\dot{\alpha}}}
\newcommand{\dbeta}{{\dot{\beta}}}

\newcommand{\N}{{\scriptscriptstyle N}}

\newcommand{\half}{{{\textstyle\frac{1}{2}}}}
\newcommand{\quarter}{{{\textstyle\frac{1}{4}}}}

\newcommand{\be}{\begin{equation} }
\newcommand{\ee}{\end{equation} }
\newcommand{\ba}{\begin{array}}
\newcommand{\ea}{\end{array}}

\newcommand{\su}{\mbox{su}}
\newcommand{\so}{\mbox{so}}
\newcommand{\SO}{\mbox{SO}}

\def\E{{\cal E}}
\def\L{{\cal L}}

\def\tr{{\rm tr}}

\def\I_M{{I_{\scriptscriptstyle M\times M}}}

\def\N{{\cal  N}}
\def\const{{\nu}}

\def\bps{{|\Psi_{\rm{BPS}}\rangle}}

\title{Supersymmetric objects in the M-theory on a pp-wave}

\author{Jeong-Hyuck Park\\
School of Physics, Korea Institute for Advanced Study\\ 207-43, Cheongryangri-Dong, Dongdaemun-Gu, Seoul 130-012, Corea\\
Electronic correspondence : \email{jhp@kias.re.kr}}

\abstract{We obtain, in a systematic way, all the classical  BPS equations  which correspond to the quantum BPS states in the  M-theory on a fully
supersymmetric pp-wave. The superalgebra of the M-theory matrix model shows that the BPS states always preserve pairs of supersymmetry, implying the
possible fractions of the unbroken supersymmetry as $\nu=2/16,\,4/16,\,6/16,\cdots$. We study their classical counterparts, and find   there are
essentially one unique set of $2/16$ BPS equations, three inequivalent types of $4/16$ BPS equations, and three inequivalent types of $8/16$ BPS
equations only, in addition to   the $16/16$ static fuzzy sphere. We discuss various supersymmetric objects as solutions. In particular, when the fuzzy
sphere rotates, the supersymmetry is further broken as $16/16\rightarrow 8/16\rightarrow 4/16$.}


\preprint{KIAS-P02052\\
hep-th/0208161}

\begin{document}

\section{Introduction}
At the present time, the most promising formalism   for the  description of  the eleven dimensional M-theory prescribes the comactification on a
light-like circle or on a small spatial circle  boosted by a large amount, $x^{-}\sim x^{-}+2\pi R$, as proposed by  Banks, Fischler, Shenker and
Susskind (BFSS) \cite{Banks:1996vh,Susskind:1997cw,Sen,Seiberg}. The sector of the theory with the discrete light cone momentum, $p_{-}=N/R$, is then
exactly described in terms of D0-brane dynamics by the BFSS matrix model or the quantum mechanics obtained by dimensionally reducing the ten dimensional
$\mbox{U}(N)$ super Yang-Mills theory. The BFSS matrix model is also in a good agreement with  the matrix  regularization of the supermembrane action in
the light cone gauge \cite{deWit,Hoppe}. However, due to the flat directions in the potential the matrix model is of continuous spectrum and has proved
very difficult to approach.\newpage

Recently \cite{Berenstein:2002jq},    Berenstein, Maldacena and Nastase (BMN)  showed that in the maximally  supersymmetric pp-wave background of the
eleven dimensional supergravity \cite{Kowalski-Glikman,Figueroa-O'Farrill1,FO2},
\begin{equation}
\ba{l}ds^{2}=-2dx^{+}dx^{-}-\Big[(\textstyle{\frac{\mu}{3}})^{2}(x_{1}^{2}+x_{2}^{2}+x_{3}^{2})
+(\textstyle{\frac{\mu}{6}})^{2}(x_{4}^{2}+\cdots+x_{9}^{2})\Big]dx^{+}dx^{+}+\displaystyle{\sum_{A=1}^{9}}\,dx^{A}dx^{A}\,,\\
{}\\
F_{123+}=\mu\,, \ea
\end{equation}
the discrete light cone quantization (DLCQ) of the M-theory  still works. With the  characteristic  mass parameter, $\mu$, the resulting new matrix
model  corresponds to a mass deformation of the BFSS matrix model without breaking any supersymmetry. Soon after, the action was rederived as a
description of the supermembrane on a pp-wave by Dasgupta, Sheikh-Jabbari and Van Raamsdonk  \cite{Dasgupta:2002hx}.\\

Thanks to   the mass parameter, the BMN  matrix model captures many interesting novel properties. The supersymmetry transformations have explicit time
dependence so that the supercharges do not commute with the Hamiltonian. As a result, the bosons and fermions have different masses. The mass terms lift
up the flat directions completely and the perturbative expansion is possible by powers of $\mu^{-1}$ \cite{Dasgupta:2002hx,Kim:2002if}.
Classical vacua are given by fuzzy spheres sitting at the origin stretching over the $1,2,3$ directions.\\

In our previous work \cite{JhpMsusy}, we studied the superalgebra of this pp-wave matrix model. We identified the superalgebra as the special unitary Lie
superalgebra, $\mbox{su}(2|4\,;2,0)$ for $\mu>0$ or $\mbox{su}(2|4\,;2,4)$ for $\mu<0$ of which  the complexification  corresponds  to $\mbox{A}(1|3)$.
After  analyzing  its root structure, we discussed the typical and atypical  representations deriving the `typicality'  condition explicitly in terms of
the energy and other four quantum numbers. In particular, we obtained the complete classification of the BPS multiplets which in general belong to a
special class of the atypical unitary representations. They are classified as   $4/16,\,8/16,\,12/16$ $\su(2)$ singlet BPS multiplets and $8/16$ $\su(4)$
singlet BPS
multiplets, in addition to the $16/16$ vacua.\\

Generically,  the   BPS state is  defined  as a state in a supermultiplet  which is  annihilated by at least  one Noether charge of the supersymmetry or
one hermitian supercharge, while  the  BPS multiplet is   defined as a unitary irreducible representation  of which either  the lowest weight or the
highest weight is a BPS state. This definition naturally leads to  a superspace with lower number of ``odd'' coordinates \cite{Ferrara:2000dv}.\\

In \cite{Keshav},    Dasgupta {\it et al.}  investigated supermultiplets which contain at least one  BPS state. These supermultiplets are not necessarily
BPS multiplets. They concluded that there can appear $2/16,4/16,6/16,8/16,12/16,16/16$ BPS states only in the supermultiplets.\newpage

Both of the   analysis above were based on the  pp-wave superalgebra which is free from the central charge.  As the central charges in the matrix models
appear as `the trace of the commutator', their absence is justified in the finite matrix models but not in the large $N$ limit.\\

In the present paper, we study the  classical counterparts of  the  quantum  BPS states.  Namely we obtain all the classical  BPS equations  which
correspond to the quantum BPS states in the  BMN matrix model. Some simple analysis of the superalgebra which may now contain  nontrivial central charges
show that the BPS states always preserve pairs of supersymmetry, implying the possible fractions of the unbroken supersymmetry as
$\nu=2/16,\,4/16,\,6/16,\cdots$.  Our main results are that    there are essentially one unique set of $2/16$ BPS equations, three inequivalent types of
$4/16$ BPS equations, and three inequivalent types of $8/16$ BPS equations only, in addition to   the $16/16$ static fuzzy sphere. We discuss various
supersymmetric objects as solutions. In particular, when the fuzzy sphere rotates on the transverse planes, the supersymmetry is further broken as
$16/16\rightarrow 8/16\rightarrow 4/16$.\\

The key tool we employ here, following \cite{jhpBPS}, is `the projection matrix' to the kernel space all the Killing spinors form. Once we are able to
write down the projection matrix in terms of the anti-symmetric products of the gamma matrices, it is straightforward to obtain the corresponding BPS
equations. In this way, the complete classification of the BPS equations in six and eight dimensions have been achieved in \cite{jhpBPS}.\\

The organization of the present paper is as follows. In section \ref{setup}, after setting up the gamma matrices and other conventions,  we review the BMN
matrix model. In particular, we write the action and the superalgebra  in a  $\mbox{so}(3)\!\times\!\mbox{so}(6)$ manifest fashion.  All the possible
central charges are included in this setup. Section \ref{susyobject} contains our main results. Firstly we discuss the general prescription to derive the
classical BPS equations using the concept of the projection matrix.  We then look at  the quantum BPS states to identify  the corresponding Killing
spinors. Sequentially  all the projection matrices  relevant to the quantum BPS states are constructed and written in terms of  the gamma matrix products.
This enables us to obtain  all the classical  BPS equations  corresponding to the quantum BPS states,  the   $2/16$ BPS equations, the three  types of
$4/16$ BPS equations, and the three  types of $8/16$ BPS equations. We discuss at least one class of solutions for each case.  In section
\ref{conclusion} we conclude  with the summary. The appendix contains the BPS equations of the ten dimensional super Yang-Mills theory as well as some
useful formulae.

\newpage

\section{M-theory matrix model on a pp-wave\label{setup}}
The action of the M-theory matrix model on a fully supersymmetric pp-wave background spells with a mass parameter, $\mu$,
\begin{equation}
{\cal S}=\displaystyle{\frac{l_{p}^{6}}{R^{3}}}\displaystyle{\int\,{\rm d}t}~ \L_{0}+\mu\L_{1}+\mu^{2}\L_{2}\,,
\end{equation}
where with $i=1,2,3$, $~a=4,5,\cdots,9$, $~A=1,2,\cdots,9$,
\begin{equation}
 \ba{l} \L_{0}=\tr\!\left(\half D_{t}X^{A}D_{t}X_{A}+\quarter [X^{A},X^{B}]^{2}+i\half
\Psi^{\dagger}D_{t}\Psi-\half\Psi^{\dagger}\Gamma^{A}[X_{A},\Psi]\right)\,,\\
{}\\
\L_{1}=i\,\tr\!\left(-\textstyle{\frac{1}{3}}\epsilon_{ijk}X^{i}X^{j}X^{k}+\textstyle{\frac{1}{8}}\Psi^{\dagger}\Gamma^{123}\Psi \right)\,,\\
{}\\
\L_{2}=-\,\half\,\tr\!\left((\textstyle{\frac{1}{3}})^{2}X^{i}X_{i}+(\textstyle{\frac{1}{6}})^{2}X^{a}X_{a}\right)\,. \ea
\end{equation}
We make a few remarks  especially compared to the original one  given by BMN \cite{Berenstein:2002jq}.  Here the Euclidean nine dimensional gamma
matrices, $\Gamma^{A}=(\Gamma^{A})^{\dagger}$,  are generic ones. Namely we do not adopt the usual real and symmetric Majorana representation.
Accordingly there exits a nontrivial   $16\times 16$ charge conjugation matrix, ${C}$,
\begin{equation}
\begin{array}{ll}
(\Gamma^{A}){}^{T}=(\Gamma^{A}){}^{\ast}={C}^{-1}\Gamma^{A}{C}\,,~~~&~~~C=C^{T}=(C^{\dagger})^{-1}\,.
\end{array}
\end{equation}
The spinor, $\Psi$,  satisfies the Majorana condition leaving  eight independent  complex components
\begin{equation}
\Psi={C}\Psi^{\ast}\,. \label{Majoranacondition}
\end{equation}
The covariant derivatives are in our convention, $D_{t}{\cal O}=\frac{d~}{dt}{\cal O}-i[A_{0},{\cal O}]$ so that $X$ and $A_{0}$ are of the mass
dimension one, while $\Psi$ has the mass dimension ${3}/{2}$. The  overall constant, ${l_{p}^{6}}/{R^{3}}$,  is set to be one henceforth.
{}\\

The supersymmetry transformations are
\be \ba{l} \delta A_{0}=i\Psi^{\dagger}\E(t)\,,~~~~~~~~~~~
\delta X^{A}=i\Psi^{\dagger}\Gamma^{A}\E(t)\,,\\
{}\\
\delta\Psi=\left(D_{t}X^{A}\Gamma_{A}-i\half [X^{A},X^{B}]\Gamma_{AB}+\mu
(-\textstyle{\frac{1}{3}}X^{i}\Gamma_{i}+\textstyle{\frac{1}{6}}X^{a}\Gamma_{a})\Gamma^{123}\right)\E(t)\,, \label{susytr}\ea\ee
where
\begin{equation}
\ba{cc} \displaystyle{\E(t)=e^{\frac{\mu}{12}\Gamma^{123}t}\E\,,}~~~&~~~\E={C}\E^{\ast}\,.\ea \label{Et}
\end{equation}
In addition there is  kinetic supersymmetry,
\begin{equation}
\ba{ccc} \delta A_{0}=\delta X^{A}=0\,, ~~~~&~~~~
\displaystyle{\delta\Psi=e^{-\frac{\mu}{4}\Gamma^{123}t}\E^{\prime}\,,}~~~&~~~\E^{\prime}={C}\E^{\prime}{}^{\ast}\,.\ea\label{Eprimet}
\end{equation}

\newpage

\subsection{Manifestation of the $\mbox{so}(3)\!\times\!\mbox{so}(6)$ structure}
To make the $\mbox{so}(3)\!\times\!\mbox{so}(6)\equiv\mbox{su}(2)\!\times\!\mbox{su}(4)$ structure of the M-theory on a pp-wave manifest,  we write the
nine dimensional gamma matrices  in terms of the three and six dimensional ones, $\sigma^{i},\gamma^{a}$, 
\begin{equation}
\begin{array}{ll}
\Gamma^{i}=\sigma^{i}\otimes\gamma^{(7)}&~~~\mbox{for~}~i=1,2,3\,,\\
{}&{}\\
\Gamma^{a}=1\otimes\gamma^{a}&~~~\mbox{for~}~a=4,5,6,7,8,9\,.
\end{array}
\end{equation}
With the choice,
\begin{equation}
\gamma^{(7)}=i\gamma^{4}\gamma^{5}\cdots\gamma^{9}=\left(\begin{array}{rr}1&0\\0&-1\end{array}\right)\,, \label{gamma7}
\end{equation}
the six dimensional gamma matrices  are in the block diagonal form
\begin{equation}
\begin{array}{cc}
\gamma^{a}=\left(\begin{array}{cc}0&\rho^{a}\\ \bar{\rho}^{a}&0\end{array}\right)\,,~~~&~~\rho^{a}\bar{\rho}^{b}+\rho^{b}\bar{\rho}^{a}=2\delta^{ab}\,.
\end{array}\label{gamma6}
\end{equation}
The fact, $\bar{\rho}^{a}=(\rho^{a})^{\dagger}$,  ensures  $\gamma^{a}$ to be hermitian. Furthermore, it is possible to set the $4\times 4$ matrices,
$\rho^{a}$, to be anti-symmetric \cite{Park:1998nr}
\begin{equation}
\begin{array}{cc} (\rho^{a})_{\dalpha\dbeta}=-(\rho^{a})_{\dbeta\dalpha}\,,~~&~~(\bar{\rho}^{a})^{\dalpha\dbeta}=-(\bar{\rho}^{a})^{\dbeta\dalpha}\,.
\end{array}\label{anti-sym}
\end{equation}
Namely six of $\rho_{a}$  form a basis of the $4\times 4$ anti-symmetric matrices. It follows that    fifteen of $\rho_{ab}=\rho_{[a}\bar{\rho}_{b]}$ and
ten of $\rho_{abc}=\rho_{[a}\bar{\rho}_{b}\rho_{c]}$ form  basis of the traceless and symmetric matrices respectively. The latter is   subject to the
self-duality,
\begin{equation}
\rho_{abc}=\frac{i}{6}\epsilon_{abcdef}\rho^{def}\,. \label{self-duality}
\end{equation}
Henceforth $\alpha,\beta=1,2$ are the ${\mbox{su}(2)}$ indices, and $\dalpha,\dbeta$ denote the ${\mbox{su}(4)}$ indices, $1,2,3,4$.\\

The nine dimensional charge conjugation matrix, ${C}$, is now explicitly
\begin{equation}
\ba{cc} {C}=\epsilon\otimes\left(\begin{array}{rr}0&-1\\1&0\end{array}\right)\,,~~~~&~~~~(\sigma^{i})^{T}=-\epsilon^{-1}\sigma^{i}\epsilon\,,\ea
\end{equation}
so that Majorana spinors contain  eight independent  complex components,
\begin{equation}
\begin{array}{cc}
\Psi=\left(\begin{array}{c} \psi_{\alpha\dalpha}\\ \\
\tilde{\psi}_{\alpha}{}^{\dalpha}\end{array}\right)\,,~~~&~~~\tilde{\psi}_{\alpha}{}^{\dalpha}=\epsilon_{\alpha\beta}(\psi^{\ast})^{\beta\dalpha}\,,\\
{}&{}\\
\E(t)=\left(\begin{array}{c} e^{i\frac{\mu}{12}t}\varepsilon_{\alpha\dalpha}\\ \\
 e^{-i\frac{\mu}{12}t}\tilde{\varepsilon}_{\alpha}{}^{\dalpha}\end{array}\right)\,,~~~&~~~
\tilde{\varepsilon}_{\alpha}{}^{\dalpha}=\epsilon_{\alpha\beta}(\varepsilon^{\ast})^{\beta\dalpha}\,.\\
\end{array}
\end{equation}
{}\\

We rewrite the Lagrangian, $\L_{0}$, $\L_{1}$, in a more $\mbox{so}(3)\!\times\!\mbox{so}(6)$ manifest fashion,
\begin{equation}
\ba{l} \L_{0}=\tr\!\left(\!\!\ba{l}\half D_{t}X^{A}D_{t}X_{A}+\quarter [X^{A},X^{B}]^{2}+i\bar{\psi}D_{t}\psi
-\bar{\psi}\sigma^{i}[X_{i},\psi]-\half\bar{\psi}\rho^{a}[X_{a},\tilde{\psi}]-\half\bar{\tilde{\psi}}\bar{\rho}^{a}[X_{a},\psi]\ea\right)\,,\\
{}\\
\L_{1}=-\tr\!\left(i\textstyle{\frac{1}{3}}\epsilon_{ijk}X^{i}X^{j}X^{k}+\textstyle{\frac{1}{4}}\bar{\psi}\psi \right)\,. \ea
\end{equation}
The supersymmetry transformation (\ref{susytr}) becomes
\begin{equation}
\ba{ccc} \delta A_{0}=i\Big(\bar{\psi}\varepsilon(t)-\bar{\varepsilon}(t)\psi\Big)\,,&\delta
X^{i}=i\Big(\bar{\psi}\sigma^{i}\varepsilon(t)-\bar{\varepsilon}(t)\sigma^{i}\psi\Big)\,,&
\delta X^{a}=i\Big(\bar{\tilde{\psi}}\bar{\rho}^{a}\varepsilon(t)-\bar{\varepsilon}(t)\rho^{a}\tilde{\psi}\Big)\,,\\
{}\\
\multicolumn{3}{l}{~~~~~\ba{ll} \delta\psi=&\!\!\Big(D_{t}X^{i}\sigma_{i}-i\half [X^{i},X^{j}]\sigma_{ij}-i\half
[X^{a},X^{b}]\rho_{ab}-i\textstyle{\frac{\mu}{3}}X^{i}\sigma_{i}\Big)\varepsilon(t)\\
{}&{}\\
{}&\,+\Big(D_{t}X^{a}\rho_{a}-i [X^{i},X^{a}]\sigma_{i}\rho_{a}-i\textstyle{\frac{\mu}{6}}X^{a}\rho_{a}\Big)\tilde{\varepsilon}(t)\,,\ea }
\ea \label{susytrso3so6}
\end{equation}
where
$\varepsilon(t)=e^{i\frac{\mu}{12}t}\varepsilon$ and $\bar{\psi}=\psi^{\dagger}$, etc.\\

The Gauss constraint reads with  $P^{A}=D_{t}X^{A}$,
\begin{equation}
i[X^{A},P_{A}]+\{\bar{\psi}^{\alpha\dalpha},\psi_{\alpha\dalpha}\}=0\,,
\end{equation}
while the equations of motion are given in the appendix (\ref{EOM}).\\

The Hamiltonian and the $\so(3)\times\so(6)$ angular momenta are explicitly
\begin{equation}
\begin{array}{ll}
H=\tr\Big(\!\!&\!\half P^{A}P_{A}-\quarter[X^{A},X^{B}]^{2}+\bar{\psi}\sigma^{i}[X_{i},\psi]
+\half\bar{\psi}\rho^{a}[X_{a},\tilde{\psi}]+\half\bar{\tilde{\psi}}\bar{\rho}^{a}[X_{a},\psi]\\
{}&{}\\
{}&+i\textstyle{\frac{\mu}{3}}\epsilon_{ijk}X^{i}X^{j}X^{k}+
\textstyle{\frac{\mu}{4}}\bar{\psi}\psi+\half(\textstyle{\frac{\mu}{3}})^{2}X^{i}X_{i}+\half(\textstyle{\frac{\mu}{6}})^{2}X^{a}X_{a}\Big)\,,
\end{array}
\end{equation}

\begin{equation}
\ba{cc} M^{ij}=\tr\Big(X^{i}P^{j}-P^{i}X^{j}-\textstyle{\frac{i}{2}}\bar{\psi}\sigma^{ij}\psi\Big)\,,~~~&~~~~
M^{ab}=\tr\Big(X^{a}P^{b}-P^{a}X^{b}-\textstyle{\frac{i}{2}}\bar{\psi}\rho^{ab}\psi\Big)\,. \ea
\end{equation}

\subsection{Supersymmetry algebra}
The Noether charge of the supersymmetry is, from (\ref{susyLtr}),  of the form
\begin{equation}
i\,\tr\left(\Psi^{\dagger}\delta\Psi\right)=\bar{\varepsilon}^{\alpha\dalpha}Q_{\alpha\dalpha}+\bar{Q}^{\alpha\dalpha}\varepsilon_{\alpha\dalpha}\,,
\label{Noether}
\end{equation}
where  the eight component supercharges are with   $\bar{Q}^{\alpha\dalpha}=(Q_{\alpha\dalpha})^{\dagger}$
\begin{equation}
\ba{ll} Q=-ie^{-i\frac{\mu}{12}t}\,\tr\Big[\Big(&(P^{i}+i\textstyle{\frac{\mu}{3}}X^{i})\sigma_{i}+i\half[X^{i},X^{j}]\sigma_{ij}
+i\half[X^{a},X^{b}]\rho_{ab}\Big)\psi\\
{}&{}\\
{}& +\Big((P^{a}-i\textstyle{\frac{\mu}{6}}X^{a})\rho_{a}+i[X^{i},X^{a}]\sigma_{i}\rho_{a}\Big)\tilde{\psi}\,\Big]\,.\ea
\end{equation}

After the standard quantization,
\begin{equation}
\ba{cc}[X^{A}{}^{l}{}_{m},P^{B}{}^{n}{}_{r}]=i\delta^{AB}\delta^{l}{}_{r}\delta_{m}{}^{n}\,,~~~&~~~~
\{\psi_{\alpha\dalpha}{}^{l}{}_{m},\bar{\psi}^{\beta\dbeta}{}^{n}{}_{r}\}=
\delta_{\alpha}{}^{\beta}\delta_{\dalpha}{}^{\dbeta}\delta^{l}{}_{r}\delta_{m}{}^{n}\,, \ea
\end{equation}
using (\ref{APP1}),  one can identify the  supersymmetry algebra of the M-theory on a fully supersymmetric pp-wave as follows up to the Gauss constraint
({\it cf.} \cite{Hyun:2002cm,Sugiyama:2002rs,Sugiyama2})
\begin{equation}
\ba{cc} [H,Q_{\alpha\dalpha}]=\textstyle{\frac{\mu}{12}} Q_{\alpha\dalpha}\,,~~~~~~~&~~~~~~
[H,\bar{Q}^{\alpha\dalpha}]=-\textstyle{\frac{\mu}{12}}\bar{Q}^{\alpha\dalpha}\,, \ea \label{HQ}
\end{equation}

\begin{equation}
\ba{ll} [M_{ij},Q_{\alpha\dalpha}]=i\textstyle{\frac{1}{2}}(\sigma_{ij})_{\alpha}{}^{\beta}Q_{\beta\dalpha}\,,~~~~&~~~~
{}[M_{ab},Q_{\alpha\dalpha}]=i\textstyle{\frac{1}{2}}(\rho_{ab})_{\dalpha}{}^{\dbeta}Q_{\alpha\dbeta}\,,\\
{}&{}\\
{}[M_{ij},\bar{Q}^{\alpha\dalpha}]=-i\textstyle{\frac{1}{2}}\bar{Q}^{\beta\dalpha}(\sigma_{ij})_{\beta}{}^{\alpha}\,,~~~~&~~~~
[M_{ab},\bar{Q}^{\alpha\dalpha}]=-i\textstyle{\frac{1}{2}}\bar{Q}^{\alpha\dbeta}(\rho_{ab})_{\dbeta}{}^{\dalpha}\,,\ea \label{MQ}
\end{equation}

\begin{equation}
\ba{cc}
[M_{i},M_{j}]=i\epsilon_{ijk}M^{k}\,,~~~~~~&~~~~~M_{i}=\textstyle{\frac{1}{2}}\epsilon_{ijk}M^{jk}\,,\\
{}&{}\\
\multicolumn{2}{c}{[M_{ab},M_{cd}]=i(\delta_{ac}M_{bd}-\delta_{ad}M_{bc}-\delta_{bc}M_{ad}+\delta_{bd}M_{ac})\,,} \ea
\end{equation}

\begin{equation}
\begin{array}{c}
{\{Q_{\alpha\dalpha},\,\bar{Q}^{\beta\dbeta}\}=2\delta_{\alpha}{}^{\beta}\delta_{\dalpha}{}^{\dbeta}H+i\textstyle{\frac{\mu}{3}}
(\sigma^{ij})_{\alpha}{}^{\beta}\delta_{\dalpha}{}^{\dbeta}M_{ij}-i\textstyle{\frac{\mu}{6}}
\delta_{\alpha}{}^{\beta}(\rho^{ab})_{\dalpha}{}^{\dbeta}M_{ab}
+\textstyle{\frac{1}{4}}(\sigma^{ij})_{\alpha}{}^{\beta}(\rho^{ab})_{\dalpha}{}^{\dbeta}R_{ij\,ab} \,,}\\
{}\\
{\{Q_{\alpha\dalpha},\,Q_{\beta\dbeta}\}=\epsilon_{\alpha\beta}(\rho^{a})_{\dalpha\dbeta}Z_{a}+
\textstyle{\frac{1}{6}}(\sigma^{i}\epsilon)_{\alpha\beta}(\rho^{abc})_{\dalpha\dbeta}Z_{iabc}\,,}\label{susyalge}
\end{array}
\end{equation}
where $R_{ij\,ab}$, $Z_{a}$, $Z_{iabc}$ are  central charges given by the `boundary terms' or the trace of the commutator. Surely they vanish for the
finite matrix models. $R_{ij\,ab}$ and $Z_{iabc}$ satisfy the reality and the anti-self-duality conditions respectively
\begin{equation}
\ba{cc} R_{ij\,ab}=-R_{ji\,ab}=-R_{ij\,ba}=(R_{ij\,ab})^{\dagger}\,,~~~&~~~Z_{iabc}=\textstyle{-\frac{i}{6}}\epsilon_{abcdef}Z_{i\,}{}^{def}\,.\ea
\end{equation}

Note that the numbers of degrees of the left and right sides in (\ref{susyalge}) match as
\begin{equation}
\ba{c}8\times 8=1+3+15+{3\times 15}\,, {}\\
{}\\
36=6+{3\times 10}\,. \ea
\end{equation}
Basically they are the decompositions of  $8\times 8$ hermitian and symmetric matrices in terms of $\sigma^{i}$ and $\rho^{a},\bar{\rho}^{b}$.\\

As shown in our previous work \cite{JhpMsusy}, in the absence of the central charges, the above superalgebra of the M-theory on a fully supersymmetric
pp-wave is identified as the special unitary Lie superalgebra, $\mbox{su}(2|4\,;2,0)$ for $\mu>0$ or $\mbox{su}(2|4\,;2,4)$ for $\mu<0$, the
complexification  of which corresponds  to $\mbox{A}(1|3)$.\\

A natural choice of the Cartan subalgebra, $\mbox{u}(1)\oplus\su(2)\oplus\su(4)$, is
\begin{equation}
\{H,~M_{12},~M_{45},~M_{67},~M_{89}\}. \label{Cartan}
\end{equation}
Any  state in  a supermultiplet or an irreducible representation of the superalgebra is specified by the  quantum numbers of the Cartan subalgebra, while
all the states in a supermultiplet carry the same  central charges.  Ref.\,\cite{JhpMsusy} contains the complete classification of the irreducible
representations of  $\mbox{A}(1|3)$.

\section{Supersymmetric objects\label{susyobject}}
Classically a bosonic configuration is  supersymmetric or BPS if there exits a nonzero constant Killing spinor, $\E$, such that the infinitesimal
supersymmetric transformation of the gaugino vanishes,
\begin{equation}
\delta\Psi=\left(D_{t}X^{A}\Gamma_{A}-i\half [X^{A},X^{B}]\Gamma_{AB}+\mu
(-\textstyle{\frac{1}{3}}X^{i}\Gamma_{i}+\textstyle{\frac{1}{6}}X^{a}\Gamma_{a})\Gamma^{123}\right)\displaystyle{e^{\frac{\mu}{12}\Gamma^{123}t}}\E=0\,.
\end{equation}
In general, a BPS configuration can have more than one Killing spinors.  The key tool we employ here, following  \cite{jhpBPS}, is the projection matrix
to the kernel space, $V$,  all the Killing spinors form.  With an orthonormal basis for the kernel, $V=\{\E_{n}|1\leq n\leq\N\}$, $\N=\mbox{dim}V\leq 16$,
the projection operator is formally
\begin{equation}
\Omega =\displaystyle{\sum_{n=1}^{\N}}~\E_{n}\E_{n}^{\dagger}\,,
\end{equation}
and satisfies\footnote{For simplicity we do not turn on the kinetic supersymmetry transformations which may cancel the dynamic supersymmetry
transformations in the large $N$ limit.}
\begin{equation}
\ba{cc}\Omega^{\dagger}=\Omega\,,~~~~&~~~~C\Omega^{\ast}C^{-1}=\Omega\,,\ea\label{twoconstraints}
\end{equation}

\begin{equation}
\Omega^{2}=\Omega\,,\label{square}
\end{equation}

\begin{equation}
\left(D_{t}X^{A}\Gamma_{A}-i\half
[X^{A},X^{B}]\Gamma_{AB}+\mu(-\textstyle{\frac{1}{3}}X^{i}\Gamma_{i}+\textstyle{\frac{1}{6}}X^{a}\Gamma_{a})\Gamma^{123}\right)
\displaystyle{e^{\frac{\mu}{12}\Gamma^{123}t}}\Omega=0\,.\label{forBPSeqns}
\end{equation}
It is worth to note that $\Omega$ is basis independent or unique for a given BPS configuration.

As the anti-symmetric products of the gamma matrices form a basis of the $16\times 16$ matrices, one can expand $\Omega$ in terms of them. Up to the
relation, $\Gamma^{12\cdots 9}=1$,   Eq.(\ref{twoconstraints}) restricts the projection matrix to be of the form
\begin{equation}
\Omega=\const\left(1+ r_{A}\Gamma^{A}+\textstyle{\frac{1}{4!}}r_{ABCD}\Gamma^{ABCD}\right)\,,
\end{equation}
where $r_{A}$ and $r_{ABCD}$ are real one and four form coefficients, while $\nu$ denotes the fraction of the unbroken supersymmetry. As the eigenvalues
of $\Omega$ are either $0$ or $1$ and the non-trivial products of the gamma matrices are traceless,
\begin{equation}
16\times\nu=\tr\Omega=\N\,.
\end{equation}

The only equation left to solve is (\ref{square}) in order to  get  the final form of the projection operator. Once it is done,   the BPS equations follow
straightforwardly from expanding (\ref{forBPSeqns}) by the anti-symmetric products of the gamma matrices and requiring each coefficient to vanish.
However, we do not know the most general solution of (\ref{square}). Unlike the cases in four, six and eight dimensions \cite{jhpBPS}, the present
isometry group, $\SO(3)\times\SO(6)$, is not big enough to reduce the number of free parameters significantly to give the essentially unique solution. It
appears there are infinitely many classical BPS equations which are not equivalent, even up to the isometry group.

However, this is a genuinely classical problem.  Once we consider the quantum aspects or the  BPS states, the classical complexity gets cleaned up and one
can identify all the projection matrices or  all the classical BPS equations relevant to the quantum BPS states.

\subsection{Quantum aspects}
The   {\it BPS state}, $\bps$, is  defined  as a state in a supermultiplet  which is  annihilated by at least  one Noether charge of the supersymmetry or
one hermitian supercharge,
\begin{equation}
\big(\bar{\varepsilon}^{\alpha\dalpha}Q_{\alpha\dalpha}+\bar{Q}^{\alpha\dalpha}\varepsilon_{\alpha\dalpha}\big)\bps=0\,.  \label{BPSH1}
\end{equation}
The corresponding sixteen component Killing spinor is
\begin{equation}
\E=\left(\begin{array}{c} \varepsilon\\ {}\\
\epsilon\varepsilon^{\ast}\end{array}\right)\,.
\end{equation}

One crucial step  we take here is to diagonalize $\Gamma^{12},\,\Gamma^{45},\,\Gamma^{67},\,\Gamma^{89}$ which are for the Cartan subalgebra we chose
(\ref{Cartan}). This is done by using the $\mbox{U}(4)$ symmetry, $\rho^{a}\rightarrow U\rho^{a}U^{T}$, $UU^{\dagger}=1$, which preserves the
anti-symmetric property of $\rho^{a}$. The appendix contains explicitly an example of  such gamma matrices (\ref{rhochoice}).\\

Now, as seen in (\ref{MQ}), each  of the sixteen supercharges  carries  definite quantum numbers of the Cartan subalgebra. In fact,\footnote{This
justifies the generality of  the  BPS equations of the ordinary ten dimensional super Yang-Mills theory we obtain  in (\ref{SYMBPS}). } any four
generators including $M_{12}$ in the Cartan subalgebra can uniquely specify all the sixteen supercharges by their quantum numbers, essentially as
$16=2^{4}$. Accordingly we have for all the $(\alpha,\dalpha)$ pairs satisfying $\varepsilon_{\alpha\dalpha}\neq 0$,
\begin{equation}
\ba{cc} Q_{\alpha\dalpha}\bps=0\,,~~~~~~~~~~~&~~~~~\bar{Q}^{\alpha\dalpha}\bps=0\,, \ea
\end{equation}
or equivalently\footnote{One can also easily check from (\ref{HQ}) that  one annihilation in (\ref{pairKilling}) implies the other.}
\begin{equation}
\ba{cc} (Q_{\alpha\dalpha}+\bar{Q}^{\alpha\dalpha})\bps=0\,,~~~~&~~~~i(Q_{\alpha\dalpha}-\bar{Q}^{\alpha\dalpha})\bps=0\,. \ea\label{pairKilling}
\end{equation}
Namely,  in the M-theory matrix model on a pp-wave, the BPS state always preserves pairs of supersymmetry, implying the possible fractions of the unbroken
supersymmetry as $\nu=2/16$, $4/16$, $6/16,\cdots,16/16$.\\

If we introduce a basis for eight component spinors, $\{\zeta^{\alpha\dalpha}\}$,  such that their $(\beta,\dbeta)$ components are
\begin{equation}
(\zeta^{\alpha\dalpha})_{\beta\dbeta}=\delta^{\alpha}_{~\beta}\delta^{\dalpha}_{~\dbeta}\,,
\end{equation}
we can write the pair of  Killing spinors for (\ref{pairKilling}),
\begin{equation}
\ba{cc} \E^{\alpha\dalpha}_{+}=\frac{1}{\sqrt{2}}\left(\begin{array}{c}\zeta^{\alpha\dalpha}\\ {}\\
\epsilon(\zeta^{\alpha\dalpha})^{\ast}\end{array}\right)\,,~~~~&~~~~\E^{\alpha\dalpha}_{-}=\frac{1}{\sqrt{2}}\left(\begin{array}{c} i\zeta^{\alpha\dalpha}\\ {}\\
-i\epsilon(\zeta^{\alpha\dalpha})^{\ast}\end{array}\right)=\Gamma^{123}\E_{+}^{\alpha\dalpha}\,.\ea\label{E+-}
\end{equation}
The corresponding $\nu=2/16$ projection matrix is then
\begin{equation}
\Omega_{\alpha\dalpha}=\E_{+}^{\alpha\dalpha}(\E_{+}^{\alpha\dalpha})^{\dagger}+\E_{-}^{\alpha\dalpha}(\E_{-}^{\alpha\dalpha})^{\dagger}
=\left(\ba{cc}\zeta^{\alpha\dalpha}(\zeta^{\alpha\dalpha})^{T}&0\\0&\epsilon\zeta^{\alpha\dalpha}(\zeta^{\alpha\dalpha})^{T}\epsilon^{-1}\ea\right)\,.
\end{equation}
From (\ref{rhochoice}) it is straightforward to expand this projection matrix in terms of the gamma matrices
\begin{equation}
\Omega_{\alpha\dalpha}=\textstyle{\frac{1}{8}}\Big(1+\lambda_{0}\Gamma^{3}
-\lambda_{1}\Gamma^{6789}-\lambda_{2}\Gamma^{8945}-\lambda_{1}\lambda_{2}\Gamma^{4567}-\lambda_{0}\lambda_{1}\Gamma^{1245}
-\lambda_{0}\lambda_{2}\Gamma^{1267}-\lambda_{0}\lambda_{1}\lambda_{2}\Gamma^{1289}\Big)\,.\label{2/16Omega}
\end{equation}
Here $\lambda_{0},\,\lambda_{1},\,\lambda_{2}$ are three independent signs,
\begin{equation}
\lambda_{0}^{2}=\lambda_{1}^{2}=\lambda_{2}^{2}=1\,,
\end{equation}
which are related to $(\alpha,\dalpha)$ or the unbroken supersymmetry, $Q_{\alpha\dalpha}+\bar{Q}^{\alpha\dalpha}$,
$\,i(Q_{\alpha\dalpha}-\bar{Q}^{\alpha\dalpha})$, as
\begin{center}
\begin{tabular}{lll|c}
$\lambda_{0}$&$\lambda_{1}$&$\lambda_{2}$&$(\alpha,\dalpha)$\\
\hline
$+$&$+$&$+$&$(1,1)$\\
$+$&$+$&$-$&$(1,2)$\\
$+$&$-$&$+$&$(1,3)$\\
$+$&$-$&$-$&$(1,4)$\\
$-$&$+$&$+$&$(2,1)$\\
$-$&$+$&$-$&$(2,2)$\\
$-$&$-$&$+$&$(2,3)$\\
$-$&$-$&$-$&$(2,4)$\\
\end{tabular}
\end{center}
This relation enables us to define $\Omega_{\lambda}\equiv\Omega_{\alpha\dalpha}$, $\lambda=(\lambda_{0},\lambda_{1},\lambda_{2})$. It is worth to note
that $\Omega_{\lambda}$'s are orthogonal to each other
\begin{equation}
\Omega_{\lambda}\Omega_{\lambda^{\prime}}=\delta_{\lambda\lambda^{\prime}}\Omega_{\lambda}\,,
\end{equation}
and  also complete
\begin{equation}
\displaystyle{\sum_{\lambda}}\,\Omega_{\lambda}=1_{16\times 16}\,.
\end{equation}

Other generic projection matrices of the fractions, $\nu=\N/16$, $\N=4,6,\cdots,16$, are then constructed by summing $\N/2$ different $2/16$ projection
operators above.
\newpage
\subsection{$2/16$ BPS configurations}
Substituting (\ref{2/16Omega}) into (\ref{forBPSeqns}) and expanding it by the anti-symmetric products of the gamma matrices, it is straightforward to
obtain the $2/16$ BPS equations,
\begin{equation}
\begin{array}{ll}
D_{t}X^{1}+i\lambda_{0}[X^{3},X^{1}]+\lambda_{0}\textstyle{\frac{\mu}{3}}X^{2}=0\,,~~~&~~~
D_{t}X^{2}+i\lambda_{0}[X^{3},X^{2}]-\lambda_{0}\textstyle{\frac{\mu}{3}}X^{1}=0\,,\\
{}&{}\\
D_{t}X^{3}=0\,,~~~&~~~\\
{}&{}\\
D_{t}X^{4}+i\lambda_{0}[X^{3},X^{4}]-\lambda_{1}\textstyle{\frac{\mu}{6}}X^{5}=0\,,~~~~&~~~~
D_{t}X^{5}+i\lambda_{0}[X^{3},X^{5}]+\lambda_{1}\textstyle{\frac{\mu}{6}}X^{4}=0\,,\\
{}&{}\\
D_{t}X^{6}+i\lambda_{0}[X^{3},X^{6}]-\lambda_{2}\textstyle{\frac{\mu}{6}}X^{7}=0\,,~~~~&~~~~
D_{t}X^{7}+i\lambda_{0}[X^{3},X^{7}]+\lambda_{2}\textstyle{\frac{\mu}{6}}X^{6}=0\,,\\
{}&{}\\
D_{t}X^{8}+i\lambda_{0}[X^{3},X^{8}]-\lambda_{1}\lambda_{2}\textstyle{\frac{\mu}{6}}X^{9}=0\,,~~~~&~~~~
D_{t}X^{9}+i\lambda_{0}[X^{3},X^{9}]+\lambda_{1}\lambda_{2}\textstyle{\frac{\mu}{6}}X^{8}=0\,,\\
{}&{}\\
{}[X^{1},X^{4}]-\lambda_{0}\lambda_{1}[X^{2},X^{5}]=0\,,~~~~&~~~~
{}[X^{1},X^{5}]+\lambda_{0}\lambda_{1}[X^{2},X^{4}]=0\,,\\
{}&{}\\
{}[X^{1},X^{6}]-\lambda_{0}\lambda_{2}[X^{2},X^{7}]=0\,,~~~~&~~~~
{}[X^{1},X^{7}]+\lambda_{0}\lambda_{2}[X^{2},X^{6}]=0\,,\\
{}&{}\\
{}[X^{1},X^{8}]-\lambda_{0}\lambda_{1}\lambda_{2}[X^{2},X^{9}]=0\,,~~~~&~~~~
{}[X^{1},X^{9}]+\lambda_{0}\lambda_{1}\lambda_{2}[X^{2},X^{8}]=0\,,\\
{}&{}\\
{}[X^{4},X^{6}]-\lambda_{1}\lambda_{2}[X^{5},X^{7}]=0\,,~~~~&~~~~
{}[X^{4},X^{7}]+\lambda_{1}\lambda_{2}[X^{5},X^{6}]=0\,,\\
{}&{}\\
{}[X^{4},X^{8}]-\lambda_{2}[X^{5},X^{9}]=0\,,~~~~&~~~~
{}[X^{4},X^{9}]+\lambda_{2}[X^{5},X^{8}]=0\,,\\
{}&{}\\
{}[X^{6},X^{8}]-\lambda_{1}[X^{7},X^{9}]=0\,,~~~~&~~~~
{}[X^{6},X^{9}]+\lambda_{1}[X^{7},X^{8}]=0\,,\\
{}&{}\\
\multicolumn{2}{c}{\lambda_{0}[X^{1},X^{2}]+\lambda_{1}[X^{4},X^{5}]+\lambda_{2}[X^{6},X^{7}]+\lambda_{1}\lambda_{2}[X^{8},X^{9}]
-i\lambda_{0}\textstyle{\frac{\mu}{3}}X^{3}=0\,.}
\end{array}
\end{equation}
In addition there exists the Gauss constraint
\begin{equation}
{}\displaystyle{\sum_{A}}\,[D_{t}X^{A},X_{A}]=0\,.
\end{equation}

Surely any solution of the  BPS equations subject to the Gauss constraint satisfies the full equations of motion (\ref{EOM}), as discussed in the
appendix.

\newpage

All the different choices for  $\lambda=(\lambda_{0},\,\lambda_{1},\,\lambda_{2})$ are $\SO(3)\times\SO(6)$ equivalent. In particular, for
$\lambda=(+,+,+)$, if we complexify the coordinates as
\begin{equation}
\begin{array}{cccc}
Z_{0}=X_{1}+iX_{2}\,,~&~Z_{1}=X_{4}+iX_{5}\,,~&~Z_{2}=X_{6}+iX_{7}\,,~&~Z_{3}=X_{8}+iX_{9}\,,
\end{array}
\end{equation}
and set $\bar{Z}_{\mu}=(Z_{\mu})^{\dagger}$, $\mu=0,1,2,3$, ~the $2/16$ BPS equations get simplified as
\begin{equation}
\begin{array}{ll}
D_{t}Z_{0}+i[X^{3},Z_{0}]-i\textstyle{\frac{\mu}{3}}Z_{0}=0\,,~~~~&~~~~D_{t}Z_{1}+i[X^{3},Z_{1}]+i\textstyle{\frac{\mu}{6}}Z_{1}=0\,,\\
{}&{}\\
D_{t}Z_{2}+i[X^{3},Z_{2}]+i\textstyle{\frac{\mu}{6}}Z_{2}=0\,,~~~~&~~~~D_{t}Z_{3}+i[X^{3},Z_{3}]+i\textstyle{\frac{\mu}{6}}Z_{3}=0\,,\\
{}&{}\\
D_{t}X^{3}=0\,,~~~~&~~~~[Z_{\mu},Z_{\nu}]=0\,,\\
{}&{}\\
\multicolumn{2}{c}{[Z_{0},\bar{Z}_{0}]+[Z_{1},\bar{Z}_{1}]+[Z_{2},\bar{Z}_{2}]+[Z_{3},\bar{Z}_{3}]-\textstyle{\frac{2}{3}}\mu X^{3}=0\,.}
\end{array}
\end{equation}
The Gauss constraint reads
\begin{equation}
{}\displaystyle{\sum_{\mu=0}^{3}}\,[D_{t}Z_{\mu},\bar{Z}_{\mu}]+[D_{t}\bar{Z}_{\mu},Z_{\mu}]=0\,.
\end{equation}
The energy is saturated, from (\ref{susyalge},~\ref{diagonalized}), by the angular momenta and the central charges,
\begin{equation}
H=\textstyle{\frac{\mu}{3}M_{12}-\frac{\mu}{6}(M_{45}+M_{67}+M_{89})+\frac{1}{2}(R_{1245}+R_{1267}+R_{1289})}\,.
\end{equation}
Solutions of the finite size surely do not carry any  central charge so that they  describe rotating supersymmetric objects. On the other hand, from the
Hodge  duality, $R_{iab}\equiv\half\epsilon_{ijk}R_{jkab}$, the nontrivial static solutions correspond  either to the longitudinal large M5 branes
stretching in the $1,2$ and at least two of $4,5,6,7,8,9$ directions or to the large membranes stretching in the third and  at least two of $4,5,6,7,8,9$
directions.\\

Though we do not know the most general solutions, it is easy to find  solutions for the D0 branes rotating in the whole space except the third direction,
\begin{equation}
\begin{array}{ll}
Z_{0}(t)=e^{i\frac{\mu}{3}t}{\cal Z}_{0}\,,~~~~&~~~~Z_{1}(t)=e^{-i\frac{\mu}{6}t}{\cal Z}_{1}\,,\\
{}&{}\\
Z_{2}(t)=e^{-i\frac{\mu}{6}t}{\cal Z}_{2}\,,~~~~&~~~~Z_{3}(t)=e^{-i\frac{\mu}{6}t}{\cal Z}_{3}\,,\\
{}&{}\\
A_{0}=X^{3}=0\,,~~~~&~~~~{\cal Z}_{\mu}~:~\mbox{diagonal~matrices}\,.
\end{array}
\end{equation}

\newpage

\subsection{$4/16$ BPS configurations - type I}
The $\nu=4/16$ projection matrices are  sum  of any two different $2/16$ projection operators. It is easy to see that  there are three inequivalent ways
of summing. The first type we consider  corresponds to the choice, $\lambda=(+++)$ and $(++-)$. The relevant $4/16$ BPS equations are
\begin{equation}
\begin{array}{ll}
D_{t}Z_{0}+i[X^{3},Z_{0}]-i\textstyle{\frac{\mu}{3}}Z_{0}=0\,,~~~~&~~~~D_{t}Z_{1}+i[X^{3},Z_{1}]+i\textstyle{\frac{\mu}{6}}Z_{1}=0\,,\\
{}&{}\\
D_{t}X^{3}=0\,,~~~~&~~~~[Z_{0},Z_{1}]=0\,,\\
{}&{}\\
X^{6}=X^{7}=X^{8}=X^{9}=0\,,~~~~&~~~~[Z_{0},\bar{Z}_{0}]+[Z_{1},\bar{Z}_{1}]-\textstyle{\frac{2}{3}}\mu X^{3}=0\,.
\end{array}
\end{equation}
The  Gauss constraint reads
\begin{equation}
{}[D_{t}Z_{0},\bar{Z}_{0}]+[D_{t}\bar{Z}_{0},Z_{0}]+[D_{t}Z_{1},\bar{Z}_{1}]+[D_{t}\bar{Z}_{1},Z_{1}]=0\,.
\end{equation}
The energy is saturated by the angular momenta as well as the central charges,
\begin{equation}
H=\textstyle{\frac{\mu}{3}M_{12}-\frac{\mu}{6}M_{45}+\frac{1}{2}R_{1245}}\,,
\end{equation}
and $M_{67}=M_{89}=R_{1267}=R_{1289}=0$.\\

Again,  without knowing  the most general solutions, we write  the D0 brane solutions rotating on the $(1,2)$ and $(4,5)$ planes,
\begin{equation}
\begin{array}{ll}
Z_{0}(t)=e^{i\frac{\mu}{3}t}{\cal Z}_{0}\,,~~~~&~~~~Z_{1}(t)=e^{-i\frac{\mu}{6}t}{\cal Z}_{1}\,,\\
{}&{}\\
A_{0}=X^{3}=X^{6}=X^{7}=X^{8}=X^{9}=0\,,~~~~&~~~~{\cal Z}_{0},~{\cal Z}_{1}~:~\mbox{diagonal~matrices}\,.
\end{array}
\end{equation}

\newpage

\subsection{$4/16$ BPS configurations - type II : $\mbox{su}(2)$ singlet, rotating fuzzy sphere}
The choice, $\lambda=(+++)$ and $(-++)$, corresponds  quantum mechanically  to the $4/16$ $\mbox{su}(2)$ singlet BPS multiplet \cite{JhpMsusy}. The energy
is saturated by the angular momenta only,
\begin{equation}
H=-\textstyle{\frac{\mu}{6}}(M_{45}+M_{67}+M_{89})\,,
\end{equation}
since $M_{12}=R_{1245}=R_{1267}=R_{1289}=0$.  The relevant  $4/16$ BPS equations are
\begin{equation}
\begin{array}{lll}
{}[X_{i},X_{j}]-i\textstyle{\frac{\mu}{3}}\epsilon_{ijk}X^{k}=0\,,~~~&~~~D_{t}X^{i}=0\,,~~~&~~~[X^{i},X^{a}]=0\,,\\
{}&{}&{}\\
D_{t}Z_{1}+i\textstyle{\frac{\mu}{6}}Z_{1}=0\,,~~~&~~~
D_{t}Z_{2}+i\textstyle{\frac{\mu}{6}}Z_{2}=0\,,~~~&~~~D_{t}Z_{3}+i\textstyle{\frac{\mu}{6}}Z_{3}=0\,,\\{}&{}&{}\\
{}[Z_{1},Z_{2}]=0\,,~~~&~~~[Z_{2},Z_{3}]=0\,,~~~&~~~[Z_{3},Z_{1}]=0\,,\\
{}&{}&{}\\
\multicolumn{3}{c}{[Z_{1},\bar{Z}_{1}]+[Z_{2},\bar{Z}_{2}]+[Z_{3},\bar{Z}_{3}]=0~~:~\mbox{Gauss~constraint\,.}}
\end{array}
\end{equation}
Note that the BPS equations themselves contain the Gauss constraint in this case.\\

For the finite matrices, the BPS equations  imply that $Z_{1}$, $Z_{2}$, $Z_{3}$ can be  simultaneously triangulized. Then the Gauss constraint tells us
that they are actually diagonal. Therefore, the above BPS equations describe the fuzzy sphere or the giant graviton rotating on the  $(4,5)$, $(6,7)$,
$(8,9)$ planes,
\begin{equation}
\begin{array}{lll}
X_{i}=\textstyle{\frac{\mu}{3}}J_{i}\,,~~~&~~~[J_{i},J_{j}]=i\epsilon_{ijk}J_{k}\,,~~~&~~~A_{0}=0\,,\\
{}&{}&{}\\
Z_{1}(t)=e^{-i\frac{\mu}{6}t}z_{1}{ 1}\,,~~~&~~~Z_{2}(t)=e^{-i\frac{\mu}{6}t}z_{2}{ 1}\,,~~~&~~~Z_{3}(t)=e^{-i\frac{\mu}{6}t}z_{3}{ 1}\,,
\end{array}
\end{equation}
where $z_{1},z_{2},z_{3}$ are arbitrary complex numbers indicating the position of the fuzzy sphere at $t=0$. This gives the most general irreducible
finite matrix solutions.\\

On the other hand, in the large $N$ limit, by setting $X^{i}=Z_{3}=A_{0}=0$   one can obtain  the rotating longitudinal flat M5 branes as found by Hyun
and Shin \cite{Hyun:2002cm}
\begin{equation}
\ba{ccc} \multicolumn{3}{c}{Z_{1}(t)=e^{-i\frac{\mu}{6}t}(x_{4}+ix_{5})\,,~~~~~~~~~Z_{2}(t)=e^{-i\frac{\mu}{6}t}(x_{6}+ix_{7})\,,}\\
{}&{}&{}\\
{}[x_{4},x_{5}]+[x_{6},x_{7}]=0\,,~&~[x_{4},x_{6}]+[x_{7},x_{5}]=0\,,~&~ [x_{4},x_{7}]+[x_{5},x_{6}]=0\,,\ea
\end{equation}
where $x_{4},x_{5},x_{6},x_{7}$ are time independent. The energy is given by $H=-\textstyle{\frac{\mu}{6}}(M_{45}+M_{67})$, and hence,  contrary to the
conventional wisdom \cite{Banks:1996nn}, the presence of the large longitudinal M5 branes do not always require the nonvanishing central charges when
they rotate.

In any case, we do not know the most general infinite matrix solutions.\newpage

\subsection{$4/16$ BPS configurations - type III}
With the choice, $\lambda=(+++)$ and $(-+-)$, the corresponding  $4/16$ BPS equations are
\begin{equation}
\begin{array}{lll}
{}[Z_{0},X^{3}]+\textstyle{\frac{\mu}{3}}Z_{0}=0\,,~&~[Z_{2},X^{3}]-\textstyle{\frac{\mu}{6}}Z_{2}=0\,,
~&~[Z_{3},X^{3}]-\textstyle{\frac{\mu}{6}}Z_{3}=0\,,\\
{}&{}&{}\\
D_{t}Z_{1}+i\textstyle{\frac{\mu}{6}}Z_{1}=0\,,~&~D_{t}X^{i}=0\,,~&~D_{t}X^{6}=D_{t}X^{7}=D_{t}X^{8}=D_{t}X^{9}=0\,,\\
{}&{}&{}\\
{}[Z_{0},Z_{2}]=0\,,~&~[Z_{0},Z_{3}]=0\,,~&~[Z_{2},Z_{3}]=0\,,\\
{}&{}&{}\\
{}[X^{4},X^{A}]=0\,,~&~[X^{5},X^{A}]=0\,,~&~[Z_{0},\bar{Z}_{0}]+[Z_{2},\bar{Z}_{2}]+[Z_{3},\bar{Z}_{3}]-\textstyle{\frac{2}{3}}\mu X^{3}=0\,.
\end{array}
\end{equation}
Note that the BPS equations themselves satisfy the Gauss constraint, $[X^{4},X^{5}]=0$, in this case too. The energy is saturated by the angular momenta
and the central charges,
\begin{equation}
H=\textstyle{-\frac{\mu}{6}M_{45}+\frac{1}{2}(R_{1267}+R_{1289})}\,,
\end{equation}
with $M_{12}=M_{67}=M_{89}=R_{1245}=0$.\\

Again we do not know the most general solutions. A particular solution we found involves a fuzzy sphere at the origin and a pair of  hyperboloids. They
are rotating on the $(4,5)$ plane. With the defining relation for a
 $\so(3)$ fuzzy sphere   and a $\so(2,1)$ hyperboloid,
\begin{equation}
\begin{array}{lll}
{}[J_{1},J_{2}]=iJ_{3}\,,~~~&~~~{}[J_{2},J_{3}]=iJ_{1}\,,~~~&~~~{}[J_{3},J_{1}]=iJ_{2}\,,\\
{}&{}&{}\\
 {}[K_{1},K_{2}]=-iK_{3}\,,~~~&~~~{}[K_{2},K_{3}]=iK_{1}\,,~~~&~~~{}[K_{3},K_{1}]=iK_{2}\,,
\end{array}\label{unprimed}
\end{equation}
our solution reads
\begin{equation}
\begin{array}{llll}
Z_{1}(t)=e^{-i\frac{\mu}{6}t}z_{1}1\,,~&~A_{0}=0\,,~&~
X_{1}=\textstyle{\frac{\mu}{3}}P_{0}J_{1}P_{0}^{\dagger}\,,~&~X_{2}=\textstyle{\frac{\mu}{3}}P_{0}J_{2}P_{0}^{\dagger}\,,\\
{}&{}&{}&{}\\
X_{6}=\textstyle{\frac{\sqrt{2}\,}{6}}\mu P_{1}K_{2}P_{1}^{\dagger}\,,~&~X_{7}=\textstyle{\frac{\sqrt{2}\,}{6}}\mu P_{1}K_{1}P_{1}^{\dagger}\,,~&~
X_{8}=\textstyle{\frac{\sqrt{2}\,}{6}}\mu P_{2}K^{\prime}_{2}P_{2}^{\dagger}\,,~&~
X_{9}=\textstyle{\frac{\sqrt{2}\,}{6}}\mu P_{2}K^{\prime}_{1}P_{2}^{\dagger}\,,\\
{}&{}\\
\multicolumn{4}{c}{ X_{3}=\textstyle{\frac{\mu}{3}}P_{0}J_{3}P_{0}^{\dagger}+\textstyle{\frac{\mu}{6}}P_{1}K_{3}P_{1}^{\dagger}
+\textstyle{\frac{\mu}{6}}P_{2}K^{\prime}_{3}P_{2}^{\dagger}\,,}
\end{array}
\end{equation}
where $P_{0}$, $P_{1}$, $P_{2}$ are projection operators to the orthogonal spaces,
\begin{equation}
\ba{ccc} P_{0}=\sum_{n}\,|3n\rangle\langle n|\,,~~&~~P_{1}=\sum_{n}\,|3n+1\rangle\langle n|\,,~~&~~P_{0}=\sum_{n}\,|3n+2\rangle\langle n|\,,\ea
\end{equation}
and $K^{\prime}_{1}$, $K^{\prime}_{2}$, $K^{\prime}_{3}$ form another $\so(2,1)$  representation   which can be different from the unprimed ones
(\ref{unprimed}). We refer \cite{so211,so212,so213,so214} for the details  of the various $\so(2,1)$ representations. The solution admits a Casimir
operator,
\begin{equation}
2X_{1}^{2}+2X_{2}^{2}+2X_{3}^{2}-X_{6}^{2}-X_{7}^{2}-X_{8}^{2}-X_{9}^{2}=\mbox{constant}\times 1\,.
\end{equation}

\newpage
\subsection{$8/16$ BPS configurations - type I : $\mbox{su}(4)$ singlet, various rotating objects}
Three distinct sets of the  $2/16$ BPS equations  are equivalent to the four distinct sets. Consequently  there exits no genuine $\nu=6/16$ classical BPS
configuration. There are three inequivalent ways of summing $2/16$ projection matrices, and hence three inequivalent sets of the $8/16$ BPS equations.\\

The first type we consider  deals with the choice, $\lambda=(+++), (++-), (+-+), (+--)$ so that quantum mechanically it corresponds to the $\mbox{su}(4)$
singlet BPS multiplet \cite{JhpMsusy}. The energy is saturated by a single angular momentum only,
\begin{equation}
H=\textstyle{\frac{\mu}{3}}M_{12}\,,
\end{equation}
and $M_{45}=M_{67}=M_{89}=R_{1245}=R_{1267}=R_{1289}=0$.

The relevant $8/16$ BPS equations are
\begin{equation}
\ba{ll}D_{t}Z_{0}+i[X^{3},Z_{0}]-i\textstyle{\frac{\mu}{3}}Z_{0}=0\,,~~~&~~~D_{t}X^{3}=0\,,\\
{}&{}\\
{}[Z_{0},\bar{Z}_{0}]-\textstyle{\frac{2}{3}}\mu X^{3}=0\,,~~~&~~~X^{4}=X^{5}=X^{6}=X^{7}=X^{8}=X^{9}=0\,,\ea
\end{equation}
while the Gauss constraint becomes
\begin{equation}
{}[Z_{0},[\bar{Z}_{0},X^{3}]]+[\bar{Z}_{0},[Z_{0},X^{3}]]=(\textstyle{\frac{2}{3}}\mu)^{2}X^{3}\,.
\end{equation}
{}\\

 Some solutions of these BPS equations have been found by Bak \cite{Bak:2002rq} and also recently by Mikhailov \cite{Mikhailov:2002wx} using different
ansatz, but the general solutions are not known. The solutions include rotating D0 branes,
\begin{equation}
\begin{array}{ll}
X^{3}=A_{0}=0\,,~~~~&~~~~Z_{0}(t)=e^{i\frac{\mu}{3}t}{\cal Z}_{0}~:~\mbox{diagonal~matrix}\,,
\end{array}
\end{equation}
rotating ellipsoidal branes with a real parameter, $\theta$,
\begin{equation}
\ba{ll}Z_{0}(t)=\textstyle{\frac{\sqrt{2}\,}{3}}\mu e^{i\frac{\mu}{3}t}(\cos\theta\,J_{1}+i\sin\theta\,J_{2})\,,~~~~&~~~~
X^{3}=A_{0}=\textstyle{\frac{\mu}{3}}\sin(2\theta) J_{3}\,,\ea
\end{equation}
rotating hyperboloids,
\begin{equation}
\ba{ll} Z_{0}(t)=\textstyle{\frac{\sqrt{2}\,}{3}}\mu e^{i\frac{\mu}{3}t}(\cosh\theta\,K_{3}+i\sinh\theta\,K_{1})\,,~~~&~~~
X^{3}=A_{0}=\textstyle{\frac{\mu}{3}}\sinh(2\theta) K_{2}\,,\ea
\end{equation}
and rotating non-spherical giant gravitons like the fuzzy torus,
\begin{equation}
\ba{lll}Z_{0}(t)=e^{i\frac{\mu}{3}t}{\cal Z}_{0}\,,~~~&~~~X_{3}=A_{0}=(\textstyle{\frac{2}{3}\mu})^{-1}[{\cal Z}_{0},\bar{\cal
Z}_{0}]\,,~~~&~~~[X^{3},{\cal Z}_{0}]=\textstyle{\frac{\mu}{3}}({\cal Z}_{0}-\theta(\bar{{\cal Z}}_{0})^{-1})\,,\\
{}&{}&{}\\
\multicolumn{3}{c}{{\cal Z}_{0}=\displaystyle{\sum_{n=1}^{N-1}}\,q_{n}|n\rangle\langle n+1|+q_{N}|N\rangle\langle 1|\,,}\\
{}&{}&{}\\
\multicolumn{3}{c}{2|q_{n}|^{2}-|q_{n+1}|^{2}-|q_{n-1}|^{2}=2(\textstyle{\frac{\mu}{3}})^{2}(1-\theta |q_{n}|^{-2})\,,~~~~~q_{0}\equiv
q_{N}\,,~~~q_{N+1}\equiv q_{1}\,.}\ea
\end{equation}

\newpage

\subsection{$8/16$ BPS configurations - type II : $\mbox{su}(2)$ singlet, rotating fuzzy sphere}
With the choice, $\lambda=(+++),(++-),(-++),(-+-)$, we deal with the $\mbox{su}(2)$ singlet BPS multiplet \cite{JhpMsusy}. The energy is saturated by a
single angular momentum only,
\begin{equation}
H=-\textstyle{\frac{\mu}{6}}M_{45}\,,
\end{equation}
and $M_{12}=M_{67}=M_{89}=R_{1245}=R_{1267}=R_{1289}=0$.

The corresponding BPS equations are
\begin{equation}
\begin{array}{lll}
{}[X_{i},X_{j}]-i\textstyle{\frac{\mu}{3}}\epsilon_{ijk}X^{k}=0\,,~~~&~~~D_{t}X^{i}=0\,,~~~&~~~D_{t}Z_{1}+i\textstyle{\frac{\mu}{6}}Z_{1}=0\,,\\
{}&{}&{}\\
X^{6}=X^{7}=X^{8}=X^{9}=0\,,~~~&~~~[X^{i},Z_{1}]=0\,,~~~&~~~[Z_{1},\bar{Z}_{1}]=0\,,
\end{array}
\end{equation}
where the last equation agrees with the Gauss constraint in this case.\\

The  solutions generically describe fuzzy spheres rotating on the $(4,5)$ plane,
\begin{equation}
\begin{array}{lll}
X_{i}=\textstyle{\frac{\mu}{3}}J_{i}\,,~~~&~~~A_{0}=0\,,~~~&~~~ Z_{1}(t)=e^{-i\frac{\mu}{6}t}z_{1}{ 1}\,.
\end{array}
\end{equation}
Of course, when the representation of the fuzzy sphere is trivial, the solutions describe the D0 branes rotating on the $(4,5)$ plane
\cite{Berenstein:2002jq}.

\subsection{$8/16$  BPS configurations - type III : static, large M2 or longitudinal M5}
With the choice, $\lambda=(+++),(++-),(--+),(---)$, we obtain the `{static}'  $4/16$ BPS equations,
\begin{equation}
\begin{array}{ll}
D_{t}X^{i}=D_{t}Z_{1}=0\,,~~~~&~~~~X^{6}=X^{7}=X^{8}=X^{9}=0\,,\\
{}&{}\\
{}[Z_{0},X^{3}]+\textstyle{\frac{\mu}{3}}Z_{0}=0\,,~~~~&~~~~[Z_{1},X^{3}]-\textstyle{\frac{\mu}{6}}Z_{1}=0\,,\\
{}&{}\\
{}[Z_{0},Z_{1}]=0\,,~~~~&~~~~[Z_{0},\bar{Z}_{0}]+[Z_{1},\bar{Z}_{1}]-\textstyle{\frac{2}{3}}\mu X^{3}=0\,.
\end{array}
\end{equation}
The Gauss constraint is trivial surely.  The energy is saturated by a  central charge only,
\begin{equation}
H=\textstyle{\frac{1}{2}}R_{1245}\,,
\end{equation}
and $M_{12}=M_{45}=M_{67}=M_{89}=R_{1267}=R_{1289}=0$. Thus, it describes  static longitudinal large M5 branes stretching in the $1,2,4,5$ directions or
static large membranes stretching in the $3,4,5$ directions.\\

A class of solutions we found involves a fuzzy sphere and a hyperboloid. The hyperboloid is stretched in the $3,4,5$ directions. With the projection
operators,
\begin{equation}
\ba{cc} P_{+}=\sum_{n}\,|2n\rangle\langle n|\,,~~~~&~~~~P_{-}=\sum_{n}\,|2n+1\rangle\langle n|\,,\ea
\end{equation}
and the gauge choice, $A_{0}=0$, our solution reads
\begin{equation}
\begin{array}{llll}
X_{1}=\textstyle{\frac{\mu}{3}}P_{+}J_{1}P_{+}^{\dagger}\,,~&~X_{2}=\textstyle{\frac{\mu}{3}}P_{+}J_{2}P_{+}^{\dagger}\,,~&~
X_{4}=\textstyle{\frac{\sqrt{2}\,}{6}}\mu P_{-}K_{2}P_{-}^{\dagger}\,,~&~X_{5}=\textstyle{\frac{\sqrt{2}\,}{6}}\mu P_{-}K_{1}P_{-}^{\dagger}\,,\\
{}&{}&{}&{}\\
\multicolumn{4}{c}{ X_{3}=\textstyle{\frac{\mu}{3}}P_{+}J_{3}P_{+}^{\dagger}+\textstyle{\frac{\mu}{6}}P_{-}K_{3}P_{-}^{\dagger}\,.}
\end{array}
\end{equation}
Again we do not know the most general solutions.

\subsection{$16/16$ BPS configurations : static fuzzy sphere}
More than four sets of the  $2/16$ BPS equations have only the static fuzzy sphere as the common solution.  Thus, there exits no genuine $\nu=10/16$,
$12/16$, $14/16$ classical BPS configuration. For the completeness we write the $16/16$ BPS equations describing the static fuzzy spheres,
\begin{equation}
\begin{array}{ccc}
[X_{i},X_{j}]-i\textstyle{\frac{\mu}{3}}\epsilon_{ijk}X^{k}=0\,,~&~D_{t}X^{i}=0\,,~&~X^{4}=X^{5}=X^{6}=X^{7}=X^{8}=X^{9}=0\,.
\end{array}
\end{equation}

\section{Conclusion\label{conclusion}}
We have obtained, in a systematic way,  all the classical  BPS equations  which correspond to the quantum BPS states in the  M-theory on a fully
supersymmetric pp-wave. \\

The superalgebra of the M-theory matrix model shows that the BPS states always preserve pairs of supersymmetry, implying the possible fractions of the
unbroken supersymmetry as $\nu=2/16,\,4/16,\,6/16,\cdots$.  Diagonalizing $\Gamma^{12},\Gamma^{45},\Gamma^{67},\Gamma^{89}$ for the Cartan subalgebra,
we  were able to identify all the   pairs of Killing spinors explicitly. There are eight of them and they are orthogonal and complete.\\

 The key tool we employed was the projection matrix to the kernel space  the Killing spinors form. The minimal, $\nu=2/16$, projection matrices were
 constructed and written in terms of  the anti-symmetric gamma matrix products. Three independent signs appearing in the expression make eight of them
 orthogonal and complete. The corresponding  $2/16$ BPS equations were then obtained from replacing the Killing spinor in the supersymmetry transformation
 formula  by the projection operator.  Expanding this formula by the gamma matrix products,   we obtained eight sets of the  $2/16$ BPS equations of different
 sign choices. Up to the isometry group, $\SO(3)\times\SO(6)$, they are all equivalent. Similarly, the BPS equations of the higher fractions,
 $\nu=\N/16$, $\N=4,6,8,\cdots$, can be obtained from the projection operator which is any $\N/2$ sum of the minimal ones.
 Effectively,  the $\N/16$  BPS equations are equivalent to the $\N/2$ sets of the $2/16$ BPS equations.\\

We found  there are essentially one unique set of $2/16$ BPS equations, three inequivalent types of $4/16$ BPS equations, and three inequivalent types of
$8/16$ BPS equations, in addition to   the $16/16$ static fuzzy sphere. In particular, three of them correspond to  the  $4/16$ $\mbox{su}(2)$,  $8/16$
$\mbox{su}(2)$ and  $8/16$ $\mbox{su}(4)$ singlet BPS multiplets found in our previous work \cite{JhpMsusy}. However, the $12/16$ $\mbox{su}(2)$ singlet
BPS multiplets do not appear as classical configurations.\\

For each BPS configuration, we obtained  the energy saturation formula in terms of the angular momenta and the central charges.
The formula contains some useful informations such as  the static properties, the rotational directions, and the stretched directions of the large
objects,  M2, M5. Our results show that    all the classical $\mbox{su}(2)$ singlet and $\mbox{su}(4)$ singlet  BPS configurations have vanishing central
charges.\newpage

According to  the superalgebra representation theory results  \cite{Keshav}, there can appear $2/16,4/16,6/16,8/16,12/16,16/16$ BPS states only in the
supermultiplets. Our results show that at the classical level, $6/16$ and $12/16$ BPS configurations  are missing.\\

In most of the cases we  were not able to obtain  the most general solutions, but  we discussed at least one class of solutions in each case.  For the
$\mbox{su}(2)$ singlet $4/16$, $8/16$ BPS equations we  obtained the  most general finite matrix solutions. They describe  the rotating fuzzy spheres.
Namely the fuzzy sphere is fully supersymmetric when it is static,  half supersymmetric when it is rotating on a single plane, $(4,5)$, quarter
supersymmetric when it is rotating on the three planes,  $(4,5),(6,7),(8,9)$,  demonstrating the supersymmetry breaking pattern as $16/16\rightarrow
8/16\rightarrow 4/16$. Some non-supersymmetric fuzzy sphere configurations have been studied in \cite{Sugiyama:2002bw}.\\

As for the D0 branes, when they rotate on the $(1,2)$ plane with the  frequency, $\mu/3$, or  $(4,5)$, $(6,7)$, $(8,9)$ planes  with the frequency,
$\mu/6$, we have the unbroken supersymmetry as
\begin{center}
\begin{tabular}{l|llll}
$~~\,\nu$&\multicolumn{4}{c}{rotating planes for D0}\\
\hline
$2/16$ & $(1,2)$ & $(4,5)$ & $(6,7)$ & $(8,9)$\\
$4/16_{\rm\,type\,I}$ & $(1,2)$ & $(4,5)$ & {} & {}\\
$4/16_{\rm\,type\,II}$ & {} & $(4,5)$ & $(6,7)$ & $(8,9)$\\
$8/16_{\rm\,type\,I}$ & $(1,2)$ & {} & {} & {}\\
$8/16_{\rm\,type\,II}$ & {} & $(4,5)$ & {} & {}\\
\end{tabular}
\end{center}
{~}\\

We found a class of solutions for the $4/16_{\rm\,type\,III}$   BPS equations which consists of a fuzzy sphere and a pair of hyperboloids rotating on the
$(4,5)$ planes. It would be interesting to find more mingled configurations which  realize the curved longitudinal M5 branes.\\

The $\mbox{su}(4)$ singlet $8/16$ BPS equations have various known solutions in the literature as rotating D0 branes, ellipsoidal branes, hyperboloids and
fuzzy torus. All of them rotate on the  $(1,2)$ plane with the frequency, $\mu/3$, having the energy, $H=\frac{\mu}{3}M_{12}$.\\

The $8/16_{\rm\,type\,III}$ BPS equations are of unique interest since they are genuinely  static equations.  They describe  static large  longitudinal
M5 branes stretching in the $1,2,4,5$ directions or static large membranes stretching in the $3,4,5$ directions. The energy is saturated by a single
central
charge, $H=\half R_{1245}$.\\

Contrary to the conventional wisdom \cite{Banks:1996nn},  the presence of the large longitudinal M5 branes do not always imply the nonvanishing  central
charges when they rotate.\\

In principle, one  could  obtain  ``$1/16,\,3/16,\,5/16,\cdots$ BPS equations'' using the projection matrix method.  The generic solution of these
equations, if any,  will correspond not to a single state in the  supermultiplets, but to a linear combination of the states, for any choice of the Cartan
subalgebra. Further investigation is required.\\

Our BPS equations are directly applicable to  the  BFSS matrix model or to the ten dimensional super Yang-Mills theory.  One simply needs to set $\mu=0$
and replace $D_{t}X_{A}$, $-i[X_{A},X_{B}]$ by $F_{0A}$, $F_{AB}$.
We present the BPS equations in  ten dimensional super Yang-Mills theory  in the appendix (\ref{SYMBPS}).\\

It would be interesting to see how the BPS configurations we obtained  will appear in the  IIA string theory on a pp-wave
\cite{Hyun:2002wu,Bonelli:2002mb} or in the DLCQ description of the longitudinal M5 branes on a pp-wave \cite{Kim:2002cr}.

\acknowledgments{The author wishes to thank  Kimyeong Lee,  Sang-Heon Yi for the useful  discussions,   Mahn-Soo Choi, Mark Van Raamsdonk for the helpful
comments, and  Nakwoo Kim, Richard Szabo  for the nice  hospitality.}

\newpage
\appendix
\section{Appendix}
Here we show that \textit{any solution of the BPS equations subject to the Gauss constraint satisfies the full equations of motion.}

First  it is useful to note that under the supersymmetry transformations (\ref{susytrso3so6}) the Lagrangian transforms as
\begin{equation}
\ba{ll} \delta\L&=\displaystyle{\tr\!\left(\delta X^{A}\frac{\partial\L}{\partial X^{A}}+\delta\dot{X}^{A}\frac{\partial\L}{\partial\dot{X}^{A}}+\delta
A_{0}\frac{\partial\L}{\partial A_{0}}+ \delta\psi_{\alpha\dalpha}\frac{\partial\L}{\partial\psi_{\alpha\dalpha}}
+\delta\dot{\psi}_{\alpha\dalpha}\frac{\partial\L}{\partial\dot{\psi}_{\alpha\dalpha}}+
\delta\bar{\psi}^{\alpha\dalpha}\frac{\partial\L}{\partial\bar{\psi}^{\alpha\dalpha}}\right)}\\
{}&{}\\
{}&=\displaystyle{\frac{d~}{dt}\tr\!\left(\delta
X^{A}\frac{\partial\L}{\partial\dot{X}^{A}}+i\delta\bar{\psi}^{\alpha\dalpha}\psi_{\alpha\dalpha}\right)}\,, \ea\label{susyLtr}
\end{equation}
from which we obtain the Noether charge of the supersymmetry (\ref{Noether}).

For the BPS solutions satisfying $\delta\psi=\delta\bar{\psi}=0$, the above relation reduces to
\begin{equation}
\displaystyle{\tr\!\left[\Psi^{\dagger}\Gamma^{A}\E(t)\left(\frac{\partial\L}{\partial
X^{A}}-\frac{d~}{dt}\left(\frac{\partial\L}{\partial\dot{X}^{A}}\right)\right) +\Psi^{\dagger}\E(t)\frac{\partial\L}{\partial A_{0}}\right]=0\,.}
\end{equation}
This equation holds for arbitrary $\Psi$, and hence
\begin{equation}
\displaystyle{\Gamma^{A}\E(t)\left.\left(\frac{\partial\L}{\partial
X^{A}}-\frac{d~}{dt}\left(\frac{\partial\L}{\partial\dot{X}^{A}}\right)\right)\right|_{\Psi=0}+\left.\E(t)\frac{\partial\L}{\partial
A_{0}}\right|_{\Psi=0}=0\,.}
\end{equation}
At this point one can safely assume  $\E(t)$ is bosonic.  Then contracting  with $\E(t)^{\dagger}\Gamma^{B}$ and using
\begin{equation}
\E(t)^{\dagger}\Gamma^{B}\Gamma^{A}\E(t)=\E(t)^{T}C^{-1}\Gamma^{B}\Gamma^{A}\E(t)=
\E(t)^{\dagger}\Gamma^{A}\Gamma^{B}\E(t)=\delta^{AB}\E(t)^{\dagger}\E(t)\,,
\end{equation}
we obtain
\begin{equation}
\displaystyle{\E(t)^{\dagger}\E(t)\left.\left(\frac{\partial\L}{\partial
X^{B}}-\frac{d~}{dt}\left(\frac{\partial\L}{\partial\dot{X}^{B}}\right)\right)\right|_{\Psi=0}
+\E(t)^{\dagger}\Gamma^{B}\E(t)\left.\frac{\partial\L}{\partial A_{0}}\right|_{\Psi=0}=0\,.}
\end{equation}
This completes our proof.\\
{}\\

The full equations of motion of the M-theory matrix model are
\begin{equation}
\ba{l} D_{t}D_{t}X_{i}+[X^{A},[X_{A},X_{i}]]+i\mu\epsilon_{ijk}X^{j}X^{k}+(\textstyle{\frac{\mu}{3}})^{2}X_{i}
-\{\bar{\psi}^{\alpha\dalpha},(\sigma_{i}\psi)_{\alpha\dalpha}\}=0\,,\\
{}\\
D_{t}D_{t}X_{a}+[X^{A},[X_{A},X_{a}]]+(\textstyle{\frac{\mu}{6}})^{2}X_{a} -\half\{\bar{\psi}^{\alpha\dalpha},(\rho_{a}\tilde{\psi})_{\alpha\dalpha}\}
-\half\{\bar{\tilde{\psi}}{}^{\alpha}{}_{\dalpha},{}(\bar{\rho}_{a}\psi)_{\alpha}{}^{\dalpha}\}=0\,,\\
{}\\
iD_{t}\psi-\textstyle{\frac{\mu}{4}}\psi-[X^{i},\sigma_{i}\psi]-[X^{a},\rho_{a}\tilde{\psi}]=0\,. \ea \label{EOM}
\end{equation}

\newpage

Our choice of the  Euclidean six dimensional gamma matrices (\ref{gamma6}) are off-block diagonal
\begin{equation}
\begin{array}{cc}
\gamma^{a}=\left(\begin{array}{cc}0&\rho^{a}\\ \bar{\rho}^{a}&0\end{array}\right)\,,~~~&~~\rho^{a}\bar{\rho}^{b}+\rho^{b}\bar{\rho}^{a}=2\delta^{ab}\,,
\end{array}
\end{equation}
where the $4\times 4$ matrices, $\rho^{a},\,\bar{\rho}^{b}$ satisfy
\begin{equation}
\ba{ccc} \bar{\rho}^{a}=(\rho^{a})^{\dagger}\,,~~~&~~~(\rho^{a})_{\dalpha\dbeta}=-(\rho^{a})_{\dbeta\dalpha}\,,
~~~&~~~(\bar{\rho}^{a})^{\dalpha\dbeta}=-(\bar{\rho}^{a})^{\dbeta\dalpha}\,. \ea
\end{equation}
Using the $\mbox{U}(4)$ symmetry, $\rho_{a}\rightarrow U\rho_{a}U^{T}$, $UU^{\dagger}=1$, which preserves the anti-symmetric property  of $\rho_{a}$, we
can diagonalize $\gamma^{45},\,\gamma^{67}\,,\gamma^{89}$ simultaneously
\begin{equation}
\ba{ccc} \rho^{45}=i\left(\ba{rrrr}1&0&0&0\\0&1&0&0\\0&0&-1&0\\0&0&0&-1\ea\right)\,,~&~
\rho^{67}=i\left(\ba{rrrr}1&0&0&0\\0&-1&0&0\\0&0&1&0\\0&0&0&-1\ea\right)\,,~&~ \rho^{89}=i\left(\ba{rrrr}1&0&0&0\\0&-1&0&0\\0&0&-1&0\\0&0&0&1\ea\right)\,.
\ea\label{diagonalized}
\end{equation}
Explicitly we have ({\it cf.} \cite{Park:1998nr})
\begin{equation}
\begin{array}{lll}
\rho^{4}=\left(\ba{cc}i\epsilon&0\\0&-i\epsilon^{-1}\ea\right)\,,~&~\rho^{5}=\left(\ba{cc}\epsilon&0\\0&\epsilon^{-1}\ea\right)\,,~&~
\rho^{6}=\left(\ba{cc}0&i\sigma^{3}\\-i(\sigma^{3})^{T}&0\ea\right)\,,\\
{}&{}&{}\\
\rho^{7}=\left(\ba{rr}0&1\\-1&0\ea\right)\,,~&~\rho^{8}=\left(\ba{cc}0&i\sigma^{1}\\-i(\sigma^{1})^{T}&0\ea\right)\,,~&~
\rho^{9}=\left(\ba{cc}0&i\sigma^{2}\\-i(\sigma^{2})^{T}&0\ea\right)\,.
\end{array}
\label{rhochoice}
\end{equation}
~\\
~\\
~\\

Evaluating the anti-commutator of the supercharges to derive the supersymmetry algebra (\ref{susyalge}), one needs the following Fierz identities for the
nine dimensional gamma matrices, $(\Gamma^{A})_{\bar{\alpha}}{}^{\bar{\beta}}$, $\bar{\alpha},\bar{\beta}=1,2,\cdots,16$,
\begin{equation}
\ba{l}
\delta^{\bar{\alpha}}{}_{\bar{\gamma}}\delta^{\bar{\beta}}{}_{\bar{\delta}}-\delta^{\bar{\alpha}}{}_{\bar{\delta}}\delta^{\bar{\beta}}{}_{\bar{\gamma}}
=\textstyle{\frac{1}{16}}(C^{-1}\Gamma^{AB})^{\bar{\alpha}\bar{\beta}}(\Gamma_{AB}C)_{\bar{\gamma}\bar{\delta}}+
\textstyle{\frac{1}{48}}(C^{-1}\Gamma^{ABC})^{\bar{\alpha}\bar{\beta}}(\Gamma_{ABC}C)_{\bar{\gamma}\bar{\delta}}\,,\\
{}\\
(\Gamma^{AB})_{\bar{\alpha}}{}^{\bar{\gamma}}(C^{-1}\Gamma_{B})^{\bar{\beta}\bar{\delta}}
+(C^{-1}\Gamma^{AB})^{\bar{\beta}\bar{\delta}}(\Gamma_{B})_{\bar{\alpha}}{}^{\bar{\gamma}}\,+\,(\bar{\gamma}\leftrightarrow\bar{\delta})=
2(\Gamma^{A})_{\bar{\alpha}}{}^{\bar{\beta}}C^{-1}{}^{\bar{\gamma}\bar{\delta}}
-2\delta_{\bar{\alpha}}{}^{\bar{\beta}}(C^{-1}\Gamma^{A})^{\bar{\gamma}\bar{\delta}}\,. \ea \label{APP1}
\end{equation}

\newpage

The $2/16$ BPS equations in  ten dimensional super Yang-Mills theory read with three independent signs,
$\lambda_{0}^{2}=\lambda_{1}^{2}=\lambda_{2}^{2}=1$,
\begin{equation}
\begin{array}{ll}
F_{0A}+\lambda_{0}F_{A3}=0\,,~~~~&~~~~\lambda_{0}F_{12}+\lambda_{1}F_{45}+\lambda_{2}F_{67}+\lambda_{1}\lambda_{2}F_{89}=0\,,\\
{}&{}\\
{}F_{14}+\lambda_{0}\lambda_{1}F_{52}=0\,,~~~~&~~~~F_{15}+\lambda_{0}\lambda_{1}F_{24}=0\,,\\
{}&{}\\
{}F_{16}+\lambda_{0}\lambda_{2}F_{72}=0\,,~~~~&~~~~F_{17}+\lambda_{0}\lambda_{2}F_{26}=0\,,\\
{}&{}\\
{}F_{18}+\lambda_{0}\lambda_{1}\lambda_{2}F_{92}=0\,,~~~~&~~~~F_{19}+\lambda_{0}\lambda_{1}\lambda_{2}F_{28}=0\,,\\
{}&{}\\
{}F_{46}+\lambda_{1}\lambda_{2}F_{75}=0\,,~~~~&~~~~F_{47}+\lambda_{1}\lambda_{2}F_{56}=0\,,\\
{}&{}\\
{}F_{48}+\lambda_{2}F_{95}=0\,,~~~~&~~~~F_{49}+\lambda_{2}F_{58}=0\,,\\
{}&{}\\
{}F_{68}+\lambda_{1}F_{97}=0\,,~~~~&~~~~F_{69}+\lambda_{1}F_{78}=0\,.
\end{array}\label{SYMBPS}
\end{equation}
In addition there exists the Gauss constraint, $D^{A}F_{0A}=0$.  The generic BPS equations of the higher fractions, $\nu=2/16,4/16,8/16,\cdots$,   are
ready to be obtained from this.

\end{document}